\documentclass[seceq]{ptptex}
\usepackage{amsmath,amssymb}

\notypesetlogo                       

\title{
Irreversible Circulation of Fluctuation and Entropy Production
}

\author{
Hiroyuki \textsc{Tomita}$^{1}$ and 
Mitsusada M. \textsc{Sano}$^{2}$
}

\inst{
$^{1}$Interdisciplinary Graduate School of Science and Engineering,\\
Kinki University,\\
Higashi-Osaka 577-8502, Japan\\

\vspace*{0.3cm}

$^{2}$Graduate School of Human and Environmental Studies, \\
Kyoto University, \\
Kyoto 606-8501, Japan
}

\abst{
Physical and chemical stochastic processes described by the master equation 
are investigated. 
The system-size expansion, called the $\Omega$-expansion, 
transforms the master equation to the corresponding Fokker-Planck equation. 
In this paper, we examine the entropy production 
for both the master equation and the corresponding Fokker-Planck equation. 
For the master equation, the exact expression of the entropy production 
was recently derived by Gaspard using Kolmogorov-Sinai entropy 
(J.Stat.Phys. \textbf{117} (2004), 599
[Errata; \textbf{126} (2006), 1109]). 
Although Gaspard's expression is derived from 
a stochastic consideration, 
it should be noted that 
it coincides with the thermodynamical expression. 
For the corresponding Fokker-Planck equation, 
by using the detailed imbalance relation,  
which appears in the process of deriving the fluctuation theorem 
through the Onsager-Machlup theory, 
the entropy production is expressed in terms of 
the {\it irreversible circulation of fluctuation},  
which was proposed by Tomita and Tomita 
(Prog.Theor.Phys. \textbf{51} (1974), 1731 
[Errata; \textbf{53} (1975), 1546b]). 
However, this expression for the corresponding Fokker-Planck equation 
differs from that of the entropy production for the master equation. 
This discrepancy is due to the difference 
between the master equation and the corresponding Fokker-Planck equation, 
namely the former treats discrete events, 
but the latter equation is an approximation of the former one.  
In fact, in the latter equation, the original discrete events are smoothed out. 
To overcome this difficulty, we propose a hypothetical 
{\it path weight principle}. 
By using this principle, 
the modified expression of the entropy production 
for the corresponding Fokker-Planck equation 
coincides with that of the master equation 
(i.e., the thermodynamical expression) 
for a simple chemical reaction system and a diffusion system.
}

\def\reaction#1#2{\stackrel{#1}{{\mathrel{\mathop{\rightleftharpoons}\limits_{#2}}}}}

\begin{document}

\maketitle


\section{Introduction}\label{sec1}
Theories on nonequilibrium systems near equilibrium have been successful. 
In fact, a number of landmarks have been achieved 
in nonequilibrium statistical physics, 
such as Onsager's reciprocal relation\cite{Onsager,dGM} 
and the Kubo formula\cite{Kubo}. 
However, there has been no satisfactory theory 
on nonequilibrium steady states (NESS) until 
the recent discovery of the fluctuation theorem focused 
on the NESS\cite{ECM,ES,GC,Kurchan,LebowitzSpohn,Maes}. 
With such developments,
the study of the NESS (i.e., the characterization of the NESS) 
is a little revival. 
The fluctuation theorem gives us some indications of how to 
investigate the NESS. 
However, theorists consider individual problems from their perspective. 
A unified viewpoint is lacking in the present study of the NESS. 

In this paper, as an attempt to improve the present situation, 
we construct a theory of the NESS 
for certain stochastic processes, i.e., the master equation and 
the Fokker-Planck equation. 
The entropy production for these equations is examined. 
For the master equation, in a pioneering work 
by Schnakenberg\cite{Schnakenberg},   
an exact expression of the entropy production 
for the master equation was derived. 
Recently Gaspard also derived this equation starting 
from Kolmogorov-Sinai (KS) entropy\cite{Gaspard}. 
His formula is 
\begin{equation}
\langle \sigma_{e} \rangle = \frac{1}{\tau} \Delta S_{i} = h^{R}-h,
\label{eq:ep_Gaspard_KS}
\end{equation}
where $h$ is the KS entropy and $h^{R}$ is the time-reversed KS entropy. 
Throughout this paper, we set the Boltzmann constant $k_{B}=1$ 
and we use the following notation for the entropy production, 
namely, the thermodynamical entropy production  
$\sigma_{e,\mbox{\scriptsize th}}$ 
and the stochastic averaged entropy production  
$\langle \sigma_{e} \rangle$. 
Equation (\ref{eq:ep_Gaspard_KS}) coincides with Schnakenberg's result. 
It should be noted that it also agrees with 
the thermodynamical result. 
Thus, the entropy production defined by the stochastic process 
coincides with the corresponding thermodynamical entropy production. 

In the 1970s, 
systematic studies were carried out on the master equation\cite{KMK,TT}. 
Kubo et al.\cite{KMK} applied the system-size expansion, 
called the $\Omega$-expansion,  
which was developed by van Kampen\cite{vanKampen}, to the master equation. 
They derived the corresponding Fokker-Planck equation and analyzed 
the behavior of the fluctuation. 
Later Tomita and Tomita developed Kubo et al.~'s work and 
emphasized the importance of the circulation of fluctuation 
in nonequilibrium states\cite{TT}. 
In particular, in the NESS, they showed that 
the probability flow circulates. 
Successively, Tomita et al. applied 
the Onsager-Machlup theory\cite{OM,MO} 
to the result of Ref.~\citen{TT}\cite{TOT}.
Unfortunately, in Refs.~\citen{TT} and \citen{TOT}, 
the entropy production was not investigated. 

Recently, Taniguchi and Cohen extended the Onsager-Machlup theory 
to several Langevin systems\cite{TC1,TC2,TC3} and 
derived a fluctuation theorem for them. 
A key relation in the fluctuation theorem is 
the {\it detailed imbalance relation} 
(they called it the nonequilibrium detailed balance relation), 
i.e., the violation of the detailed balance relation. 
In this paper, on the basis of this key relation, 
we shall evaluate the entropy production of 
the Fokker-Planck equation derived from the master equation. 
However, this entropy production does not coincide with the entropy production 
for the original master equation. 
The reason for this will be examined in detail. 
The difference between them is due to the fact that 
our master equation treats discrete events, 
but the Fokker-Planck equation is an approximation of it. 
In the Fokker-Planck equation, the original discrete events 
are smoothed out. 
To overcome this difficulty, 
we propose a {\it path weight principle}. 
Using this path weight principle, 
the modified entropy production for the corresponding Fokker-Planck equation 
coincides with that of the original master equation, 
which is simply the thermodynamical result. 

The organization of this paper is as follows. 
In \S~\ref{sec2}, 
the master equation is introduced and 
the expression for its entropy production is given. 
After that, the master equation 
is transformed to the corresponding Fokker-Planck equation 
by the $\Omega$-expansion. 
In \S~\ref{sec3}, 
the Onsager-Machlup theory is applied to 
the corresponding Fokker-Planck equation. 
By calculating the path probability, 
the detailed imbalance relation is derived. 
The entropy production term is also determined. 
In \S~\ref{sec4}, 
two example cases are considered. 
One is a chemical reaction network.
The other is a one-dimensional diffusion system. 
For both cases, the entropy production disagrees with 
the thermodynamical result. 
In \S~\ref{sec5}, 
to improve the result of the previous section, 
the {\it path weight principle} is proposed and applied 
to the two examples. 
As a result, 
the modified entropy production based on the path weight principle 
coincides with the thermodynamical result. 
In \S~\ref{sec6}, we summarize the results. 

\section{Master equation, $\Omega$-expansion, 
and Fokker-Planck equation}\label{sec2}
In this section, we review Gaspard's result\cite{Gaspard} and 
the results of the 1970s\cite{vanKampen,KMK,TT,TOT}. 

Our starting point is the master equation. 
The master equation describes physical and chemical processes, 
such as diffusion systems and chemical reaction networks. 
In \S~\ref{sec4}, we give two such examples.  
The master equation is given by 
\begin{equation}
\frac{\partial}{\partial t} P(\boldsymbol{X};t) = 
- \sum_{\boldsymbol{X}'} W(\boldsymbol{X}\rightarrow \boldsymbol{X}') 
P (\boldsymbol{X};t) 
+ \sum_{\boldsymbol{X}'} W(\boldsymbol{X}'\rightarrow \boldsymbol{X}) 
P(\boldsymbol{X}';t),
\label{eq:master_eq}
\end{equation}
where $\boldsymbol{X}=(X_{1},X_{2},\dots,X_{N})^{t}$ is the variable of the state. 
$P(\boldsymbol{X};t)$ is the probability distribution 
that the system is in state $\boldsymbol{X}$ at time $t$. 
$W(\boldsymbol{X} \rightarrow \boldsymbol{X}')$ is the transition probability rate, 
i.e., the probability that the system performs a transition 
from state $\boldsymbol{X}$ to state $\boldsymbol{X}'$ in a unit time. 
The entropy production for this master equation was recently calculated 
by Gaspard using KS entropy\cite{Gaspard}.
Its expression is given by 
\begin{equation}
\langle \sigma_{e} \rangle = 
\frac{1}{2} \sum_{\boldsymbol{X},\boldsymbol{X}'}
\left \{ P^{\mbox{\scriptsize st}}(\boldsymbol{X}) W(\boldsymbol{X}\rightarrow \boldsymbol{X}') - 
P^{\mbox{\scriptsize st}}(\boldsymbol{X}') W(\boldsymbol{X}'\rightarrow \boldsymbol{X}) 
\right \} 
\ln \frac{P^{\mbox{\scriptsize st}}(\boldsymbol{X}) W(\boldsymbol{X}\rightarrow \boldsymbol{X}') }
{P^{\mbox{\scriptsize st}}(\boldsymbol{X}') W(\boldsymbol{X}'\rightarrow \boldsymbol{X}) },
\label{eq:ep_Gaspard_exact}
\end{equation}
where $P^{\mbox{\scriptsize st}}(\boldsymbol{X})$ is 
the probability distribution for the NESS. 
This expression is obtained by rewriting Eq.~(\ref{eq:ep_Gaspard_KS}) 
and is equivalent to the expression originally 
obtained by Schnakenberg\cite{Schnakenberg}. 
Thus, in this paper, we call Eq.~(\ref{eq:ep_Gaspard_exact}) 
the Schnakenberg-Gaspard expression. 
In addition, it should be noted 
that this expression is simply the thermodynamical expression. 
For chemical reaction systems, 
Eq.~(\ref{eq:ep_Gaspard_exact}) is rewritten in the form of 
the sum of the products of the reaction rate and the affinity. 
Thus, the stochastic consideration gives the thermodynamical result 
for this problem. 

Here the connection between the Schnakenberg-Gaspard expression, 
i.e., Eq.~(\ref{eq:ep_Gaspard_exact}),  
and the fluctuation theorem (i.e., the path probability ratio) is shown. 
The path probability ratio between the forward path and 
the reverse path is given by
\begin{equation}
\frac{P^{\mbox{\scriptsize st}}(\boldsymbol{A})W_{\mbox{\scriptsize path}}(\boldsymbol{A}\rightarrow \boldsymbol{B})}
{P^{\mbox{\scriptsize st}}(\boldsymbol{B})W_{\mbox{\scriptsize path}}(\boldsymbol{B}\rightarrow \boldsymbol{A})} 
= \exp [ \Sigma(\boldsymbol{A}\rightarrow \boldsymbol{B}) ],
\label{eq:dir}
\end{equation}
where $\Sigma(\boldsymbol{A}\rightarrow \boldsymbol{B})$ is the entropy production 
for the path $\boldsymbol{A}\rightarrow \boldsymbol{B}$. 
This relation is a key relation in the derivation of the fluctuation theorem. 
If the time-reversal symmetry is satisfied, then the right-hand side of 
Eq.~(\ref{eq:dir}) is $1$. 
Now we assume that the forward path is given by 
\begin{equation}
\boldsymbol{A} \rightarrow \boldsymbol{X}^{(1)} \rightarrow \boldsymbol{X}^{(2)} 
\rightarrow \dots \rightarrow \boldsymbol{X}^{(T-1)} \rightarrow \boldsymbol{B}. 
\end{equation}
The path probability is given by the step-by-step transition probabilities as
\begin{equation}
W_{\mbox{\scriptsize path}}(\boldsymbol{A}\rightarrow \boldsymbol{B}) 
= W(\boldsymbol{A}\rightarrow \boldsymbol{X}^{(1)})
W(\boldsymbol{X}^{(1)}\rightarrow \boldsymbol{X}^{(2)})\cdots 
W(\boldsymbol{X}^{(T-1)}\rightarrow \boldsymbol{B}),
\end{equation}
namely in a Markov chain. 
Let the transition probability be
\begin{equation}
W^{(t)}(\boldsymbol{A}\rightarrow \boldsymbol{X}') 
= \sum_{\boldsymbol{X}^{(1)}}\sum_{\boldsymbol{X}^{(2)}}\cdots \sum_{\boldsymbol{X}^{(t-1)}} 
W_{\mbox{\scriptsize path}}(\boldsymbol{A}\rightarrow \boldsymbol{X}').
\end{equation}
The path probability ratio between the forward and reverse paths 
is given by 
\begin{eqnarray}
& & \log 
\frac{P^{\mbox{\scriptsize st}}(\boldsymbol{A})W_{\mbox{\scriptsize path}}(\boldsymbol{A}\rightarrow \boldsymbol{B})}
{P^{\mbox{\scriptsize st}}(\boldsymbol{B})W_{\mbox{\scriptsize path}}(\boldsymbol{B}\rightarrow \boldsymbol{A})} \nonumber \\
& = & 
\sum_{t=0}^{T-1} \log
\frac{P^{\mbox{\scriptsize st}}(\boldsymbol{X}^{(t)})W(\boldsymbol{X}^{(t)}\rightarrow \boldsymbol{X}^{(t+1)})}
{P^{\mbox{\scriptsize st}}(\boldsymbol{X}^{(T-t)})W(\boldsymbol{X}^{(T-t)}\rightarrow \boldsymbol{X}^{(T-t-1)})} 
\nonumber \\
& = & 
\log \frac{P^{\mbox{\scriptsize st}}(\boldsymbol{A})W(\boldsymbol{A}\rightarrow \boldsymbol{X}^{(1)})}
{P^{\mbox{\scriptsize st}}(\boldsymbol{X}^{(1)})W(\boldsymbol{X}^{(1)}\rightarrow \boldsymbol{A})}
+ 
\log \frac{P^{\mbox{\scriptsize st}}(\boldsymbol{X}^{(T-1)})
W(\boldsymbol{X}^{(T-1)}\rightarrow \boldsymbol{B})}
{P^{\mbox{\scriptsize st}}(\boldsymbol{B})W(\boldsymbol{B}\rightarrow \boldsymbol{X}^{(T-1)})}
\nonumber \\
&  &
+\sum_{t=1}^{T-2}
\log
\frac{P^{\mbox{\scriptsize st}}(\boldsymbol{X}^{(t)})W(\boldsymbol{X}^{(t)}\rightarrow \boldsymbol{X}^{(t+1)})}
{P^{\mbox{\scriptsize st}}(\boldsymbol{X}^{(t+1)})W(\boldsymbol{X}^{(t+1)}\rightarrow \boldsymbol{X}^{(t)})}. 
\end{eqnarray}
Before taking the limit $T\rightarrow \infty$, the path average is obtained. 
Then we have
\begin{eqnarray}
\langle \Sigma (\boldsymbol{A}\rightarrow \boldsymbol{B}) \rangle 
& = & \mbox{(both end terms)} \nonumber \\
& & + \sum_{t=1}^{T-2} \sum_{\mbox{\scriptsize all paths}}
P^{\mbox{\scriptsize st}}(\boldsymbol{A})W^{(t)}(\boldsymbol{A}\rightarrow \boldsymbol{X}^{(t)})
W(\boldsymbol{X}^{(t)}\rightarrow \boldsymbol{X}^{(t+1)})
\nonumber \\
& & 
\times W(\boldsymbol{X}^{(t+1)}\rightarrow \boldsymbol{B})
\log
\frac{P^{\mbox{\scriptsize st}}(\boldsymbol{X}^{(t)})W(\boldsymbol{X}^{(t)}\rightarrow \boldsymbol{X}^{(t+1)})}
{P^{\mbox{\scriptsize st}}(\boldsymbol{X}^{(t+1)})W(\boldsymbol{X}^{(t+1)}\rightarrow \boldsymbol{X}^{(t)})}. 
\end{eqnarray}
For the NESS, we assume that 
\begin{equation}
\sum_{\boldsymbol{A}} P^{\mbox{\scriptsize st}}(\boldsymbol{A}) 
W^{(t)}(\boldsymbol{A} \rightarrow \boldsymbol{X}) 
= P^{\mbox{\scriptsize st}}(\boldsymbol{X}). 
\end{equation}
We use the Bayes relation  
\begin{equation}
W(\boldsymbol{X}^{(t+1)}\rightarrow \boldsymbol{B}) = 
P^{\mbox{\scriptsize st}}(\boldsymbol{X}^{(t+1)})^{-1} 
\overline{W}(\boldsymbol{X}^{(t+1)}\rightarrow \boldsymbol{B})
P^{\mbox{\scriptsize st}}(\boldsymbol{B}),
\end{equation}
and
\begin{equation}
\sum_{\boldsymbol{B}}\overline{W}(\boldsymbol{X}^{(t+1)}\rightarrow \boldsymbol{B}) 
P^{\mbox{\scriptsize st}}(\boldsymbol{B})
= P^{\mbox{\scriptsize st}}(\boldsymbol{X}^{(t+1)}), 
\end{equation}
where the bar denotes the destined conditional probability. 
Thus, the average entropy production is given by 
\begin{eqnarray}
\langle \sigma_{e} \rangle & = & 
\lim_{T\rightarrow \infty} \frac{1}{T} 
\langle \Sigma (\boldsymbol{A}\rightarrow \boldsymbol{B}) \rangle \nonumber \\
& = & 
\sum_{\boldsymbol{X},\boldsymbol{X}'} P^{\mbox{\scriptsize st}}(\boldsymbol{X})
W(\boldsymbol{X}\rightarrow \boldsymbol{X}') 
\log
\frac{P^{\mbox{\scriptsize st}}(\boldsymbol{X})W(\boldsymbol{X}\rightarrow \boldsymbol{X}')}
{P^{\mbox{\scriptsize st}}(\boldsymbol{X}')W(\boldsymbol{X}'\rightarrow \boldsymbol{X})}.
\label{eq:SG_relation}
\end{eqnarray}
Equation (\ref{eq:SG_relation}) is simply 
the Schnakenberg-Gaspard expression, 
i.e., Eq.~(\ref{eq:ep_Gaspard_exact}). 
Throughout this paper, we call this relation 
the detailed imbalance expression of entropy production. 
The fluctuation theorem involves the asymptotic behavior 
of long-time fluctuation. 
However, note that the entropy production in the NESS is determined 
by the detailed imbalance of short-time fluctuation 
in this formula. 

As in Refs.~\citen{KMK} and ~\citen{vanKampen}, 
the $\Omega$-expansion is used for Eq.~(\ref{eq:master_eq}). 
Here $\Omega$ is a variable that is related to the system size. 
For chemical reaction systems, $\Omega$ should be on the order of 
the Avogadro number or the volume of the system. 
We set 
\begin{equation}
W(\boldsymbol{X}\rightarrow \boldsymbol{X}+\boldsymbol{r}) 
= \Omega \; w \left ( \frac{\boldsymbol{X}}{\Omega};\boldsymbol{r}\right )
\end{equation}
and scale the variable $\boldsymbol{X}$,  
\begin{equation}
\boldsymbol{x} = \frac{\boldsymbol{X}}{\Omega}. 
\end{equation}
The master equation is rewritten as 
\begin{equation}
\frac{\partial }{\partial t} p(\boldsymbol{x};t) 
= - \Omega \sum_{\boldsymbol{r}} 
(1-e^{-\epsilon \boldsymbol{r}\cdot\frac{\partial}{\partial \boldsymbol{x}}})
w(\boldsymbol{x};\boldsymbol{r}) p(\boldsymbol{x};t),
\label{eq:KM}
\end{equation}
where $\epsilon = 1/\Omega$ is used. 
The probability distribution is now scaled as 
\begin{equation}
\Omega^{N} P(\boldsymbol{X};t) = p(\boldsymbol{x};t).
\end{equation}
Expanding the right-hand side of Eq.~(\ref{eq:KM}) in terms of $\epsilon$,  
we obtain the following equation:  
\begin{equation}
\frac{\partial }{\partial t} p(\boldsymbol{x};t) 
= \sum_{n=1}^{\infty}\frac{\epsilon^{n-1}}{n!}
\left ( -\frac{\partial }{\partial \boldsymbol{x}} \right )^{n} 
\cdot \boldsymbol{c}_{n}(\boldsymbol{x})p(\boldsymbol{x};t),
\label{eq:Omega_expansion}
\end{equation}
where 
\begin{equation}
\boldsymbol{c}_{n}(\boldsymbol{x}) = 
\sum_{\boldsymbol{r}} (\boldsymbol{r})^{n} w(\boldsymbol{x};\boldsymbol{r})
\end{equation}
is the $n$th moment of the transition probability rate. 
In particular, $\boldsymbol{c}_{1}(\boldsymbol{x})$ and 
$\boldsymbol{c}_{2}(\boldsymbol{x})$ are given by 
\begin{equation}
\boldsymbol{c}_{1,k}(\boldsymbol{x})  =  
\sum_{\boldsymbol{r}} r_{k} \;w(\boldsymbol{x};\boldsymbol{r}), \; 
\boldsymbol{c}_{2,kl}(\boldsymbol{x})  =  
\sum_{\boldsymbol{r}} r_{k} r_{l} \;w(\boldsymbol{x};\boldsymbol{r}).
\label{eq:moments}
\end{equation}
By considering the lowest order of $\epsilon$, 
we obtain the following equation:
\begin{equation}
\frac{\partial }{\partial t} p(\boldsymbol{x};t) 
= - \frac{\partial }{\partial \boldsymbol{x}}\cdot \boldsymbol{c}_{1}(\boldsymbol{x}) p(\boldsymbol{x};t)
+ \frac{\epsilon}{2} \frac{\partial }{\partial \boldsymbol{x}}
\frac{\partial }{\partial \boldsymbol{x}}\cdot \boldsymbol{c}_{2}(\boldsymbol{x}) p(\boldsymbol{x};t).
\end{equation}
Here the $\boldsymbol{x}$-dependence of $\boldsymbol{c}_{2}(\boldsymbol{x})$ 
is explicitly considered. Then we obtain 
\begin{equation}
\frac{\partial }{\partial t} p(\boldsymbol{x};t) 
= - \frac{\partial }{\partial \boldsymbol{x}}\cdot 
\left [ 
\boldsymbol{c}_{1}(\boldsymbol{x}) - \frac{\epsilon}{2} \boldsymbol{h}(\boldsymbol{x} )
- \frac{\epsilon}{2} \boldsymbol{c}_{2}(\boldsymbol{x})
\cdot \frac{\partial }{\partial \boldsymbol{x}}
\right ]
p(\boldsymbol{x};t),
\end{equation}
where 
\begin{equation}
\boldsymbol{h}(\boldsymbol{x}) = \boldsymbol{c}_{2}(\boldsymbol{x}) \cdot 
\stackrel{\leftarrow}{\frac{\partial }{\partial \boldsymbol{x}}}.
\end{equation}
Here the term $\epsilon \boldsymbol{h}(\boldsymbol{x})/2$ can be neglected, 
because compared with $\boldsymbol{c}_{1}$, it is on the order of $\epsilon$. 
For later use,\footnote{
Here we used the standard definition of $\boldsymbol{D}$, 
which is different from that in Ref.~\citen{TT} by a factor of $1/2$.} we set 
\begin{eqnarray}
\boldsymbol{D}(\boldsymbol{x}) & = & \frac{1}{2}\boldsymbol{c}_{2}(\boldsymbol{x}),
\label{eq:def_D}
\end{eqnarray}
which is the diffusion-constant matrix in the probability space. 
If the fluctuation is assumed to be normal\cite{KMK}, i.e., 
\begin{equation}
p(\boldsymbol{x};t) =  A(\boldsymbol{x};t) e^{-\Omega \phi(\boldsymbol{x};t)},
\end{equation}
the probability distribution is well approximated 
as a Gaussian distribution on the order of 
$\epsilon$, 
\begin{equation}
\phi(\boldsymbol{x};t)  =  \frac{1}{2}(\boldsymbol{x}-\boldsymbol{y}(t))^{t}\cdot 
\boldsymbol{\sigma}^{-1}
\cdot (\boldsymbol{x}-\boldsymbol{y}(t)).
\end{equation}
Then we obtain the time evolution of $\boldsymbol{y}(t)$ and $\boldsymbol{\sigma}(t)$ 
as follows:  
\begin{eqnarray}
\frac{d\boldsymbol{y}}{dt} & = & \boldsymbol{c}_{1}(\boldsymbol{y}),\label{eq:yt}\\
\frac{d\boldsymbol{\sigma}}{dt} & = & \boldsymbol{K}\cdot \boldsymbol{\sigma} + 
\boldsymbol{\sigma}\cdot\widetilde{\boldsymbol{K}}
+ \boldsymbol{c}_{2}(\boldsymbol{y}),
\label{eq:time_evol_sigma}
\end{eqnarray}
where 
\begin{equation}
K_{kl}(\boldsymbol{y}) = \frac{\partial c_{1,k}(\boldsymbol{y})}{\partial y_{l}}.
\label{eq:def_K}
\end{equation}
The matrix elements of $\boldsymbol{\sigma}$ are given by 
\begin{equation}
\sigma_{ij} = \int d\boldsymbol{\xi} \; 
\xi_{i} \xi_{j} \; \tilde{p}(\boldsymbol{\xi};t),
\end{equation}
where 
\begin{equation}
\boldsymbol{x} = \boldsymbol{y}(t) + \epsilon^{1/2} \boldsymbol{\xi} 
\end{equation}
and 
$\tilde{p}(\boldsymbol{\xi};t)$ is the probability distribution 
for $\boldsymbol{\xi}$. 
Thus, $\boldsymbol{\sigma}$ and $\boldsymbol{g}$ correspond  
to the definitions in Ref.~\citen{TT}.
Solving Eq.~(\ref{eq:yt}) and inserting its solution into $\boldsymbol{D}(\boldsymbol{x})$, 
we obtain $\boldsymbol{y}(t)$ and $\boldsymbol{D}(t)$. 
Therefore, as a result, we obtain the Fokker-Planck equation: 
\begin{equation}
\frac{\partial }{\partial t} p(\boldsymbol{x};t) 
= - \frac{\partial }{\partial \boldsymbol{x}}\cdot 
\left [ 
\boldsymbol{c}_{1}(\boldsymbol{x}) 
- \epsilon \boldsymbol{D}(t) \cdot \frac{\partial }{\partial \boldsymbol{x}}
\right ]
p(\boldsymbol{x};t).
\label{eq:Fokker_Planck_eq}
\end{equation}
Hereafter we investigate this Fokker-Planck equation in detail.

From Eq.~(\ref{eq:Fokker_Planck_eq}), 
several interesting properties for the NESS have been discovered\cite{TT}. 
The probability distribution for the NESS is given by 
\begin{equation}
p^{\mbox{\scriptsize st}}(\boldsymbol{x}) = 
\frac{1}
{\sqrt{(2\pi \epsilon)^{N}\det(\boldsymbol{\sigma}^{\mbox{\scriptsize st}})}}
\exp[-\Omega \phi(\boldsymbol{x})],
\end{equation}
where 
\begin{equation}
\phi(\boldsymbol{x}) = 
\frac{1}{2} (\boldsymbol{x}-\langle \boldsymbol{x} \rangle)^{t} 
\cdot \boldsymbol{g}^{\mbox{\scriptsize st}} \cdot 
(\boldsymbol{x}-\langle \boldsymbol{x} \rangle)
\end{equation}
and 
\begin{equation}
\boldsymbol{g}^{\mbox{\scriptsize st}} = (\boldsymbol{\sigma}^{\mbox{\scriptsize st}})^{-1}.
\end{equation}
$\boldsymbol{\sigma}^{\mbox{\scriptsize st}}$ is 
the variance matrix of the probability distribution 
of the NESS. 
In this case, the phenomenological equation is given by 
\begin{eqnarray}
\boldsymbol{X} & = & - \frac{\partial \phi}{ \partial \boldsymbol{x}}  \nonumber \\
& = & -\boldsymbol{g}^{\mbox{\scriptsize st}}\cdot (\boldsymbol{x} -\langle \boldsymbol{x}\rangle).
\end{eqnarray}
If $\boldsymbol{c}_{1}(\boldsymbol{x})$ is given as 
\begin{equation}
\boldsymbol{c}_{1}(\boldsymbol{x}) = \boldsymbol{K}^{\mbox{\scriptsize st}}\cdot \boldsymbol{x} + \boldsymbol{c},
\label{eq:c1}
\end{equation}
(two examples in \S~\ref{sec4} are discussed for this case), 
we have
\begin{eqnarray}
\dot{\boldsymbol{x}} & = & \boldsymbol{K}^{\mbox{\scriptsize st}}\cdot \boldsymbol{x} + \boldsymbol{c} \nonumber \\
& = & \boldsymbol{K}^{\mbox{\scriptsize st}} \cdot (\boldsymbol{x}-\langle \boldsymbol{x} \rangle )\nonumber \\
& = & -\boldsymbol{L}\cdot \boldsymbol{X}.
\end{eqnarray}
Here we have used the fact that $\dot{\boldsymbol{x}}=0 \Rightarrow 
\langle \boldsymbol{x} \rangle = -(\boldsymbol{K}^{\mbox{\scriptsize st}})^{-1}\cdot \boldsymbol{c}$, 
where $\boldsymbol{L}$ is the Onsager coefficient. 
Also we have the following relations\cite{TT}, 
\begin{eqnarray}
\boldsymbol{L} & = & -\boldsymbol{K}^{\mbox{\scriptsize st}}(\boldsymbol{g}^{\mbox{\scriptsize st}})^{-1} = \boldsymbol{D}^{\mbox{\scriptsize st}} + \boldsymbol{\alpha}, \\
\boldsymbol{\alpha} & = & - \boldsymbol{K}^{\mbox{\scriptsize st}} \boldsymbol{\sigma}^{\mbox{\scriptsize st}} - \boldsymbol{D}^{\mbox{\scriptsize st}} \nonumber \\
& = & \frac{1}{2} (\boldsymbol{\sigma}^{\mbox{\scriptsize st}}\widetilde{\boldsymbol{K}}^{\mbox{\scriptsize st}}-\boldsymbol{K}^{\mbox{\scriptsize st}} \boldsymbol{\sigma}^{\mbox{\scriptsize st}}).
\label{eq:def_alpha}
\end{eqnarray}
The matrix $\boldsymbol{\alpha}$ vanishes for the case 
that the detailed balance is satisfied, i.e., in equilibrium. 
However, for the NESS, $\boldsymbol{\alpha}$ is non zero in general. 
In addition, $\boldsymbol{\alpha}$ is an antisymmetric matrix. 
As shown in Ref.~\citen{TT}, $\boldsymbol{\alpha}$ is 
a measure of the circulation of fluctuation. 
Therefore, $\boldsymbol{\alpha}$ is called 
the {\it irreversible circulation of fluctuation}. 

Now we set 
\begin{equation}
\boldsymbol{v}(\boldsymbol{x}) = \dot{\boldsymbol{y}}(\boldsymbol{x}) 
+ \boldsymbol{D}^{\mbox{\scriptsize st}}\cdot \nabla \phi(\boldsymbol{x}).
\end{equation}
We call the vector $\boldsymbol{v}$ the {\it irreversible circulation velocity}. 
If $\dot{\boldsymbol{y}}(\boldsymbol{x})=\boldsymbol{K}^{\mbox{\scriptsize st}}\cdot \boldsymbol{x} + \boldsymbol{c}$, 
the irreversible circulation velocity can be rewritten as 
\begin{eqnarray}
\boldsymbol{v}(\boldsymbol{x}) & =& \boldsymbol{K}^{\mbox{\scriptsize st}}\cdot(\boldsymbol{x}-\langle \boldsymbol{x} \rangle) + 
\boldsymbol{D}^{\mbox{\scriptsize st}}\cdot \nabla \phi(\boldsymbol{x}) \nonumber \\
& = & (\boldsymbol{K}^{\mbox{\scriptsize st}}+ 
\boldsymbol{D}^{\mbox{\scriptsize st}}\boldsymbol{g}^{\mbox{\scriptsize st}}) 
\cdot(\boldsymbol{x}-\langle \boldsymbol{x} \rangle) \nonumber \\
& = & -\boldsymbol{\alpha} \cdot \nabla \phi(\boldsymbol{x}) \nonumber \\
& = & \boldsymbol{\alpha}\cdot \boldsymbol{X}(\boldsymbol{x}).
\end{eqnarray}
Therefore, we have 
\begin{equation}
\boldsymbol{v}(\boldsymbol{x}) = \boldsymbol{j}(\boldsymbol{x}) / P^{\mbox{\scriptsize st}}(\boldsymbol{x}). 
\end{equation}
Here $\boldsymbol{j}(\boldsymbol{x})$ is the current of the probability. 
This relation was also obtained in Ref.~\citen{Seifert}.

In addition, the irreversible circulation $\boldsymbol{\alpha}$ is related to 
the breaking of the fluctuation-dissipation theorem. 
If the fluctuation-dissipation theorem is satisfied, 
i.e., $\boldsymbol{\alpha}=0$, then  
\begin{equation}
-\boldsymbol{K}^{\mbox{\scriptsize st}} \boldsymbol{\sigma}^{\mbox{\scriptsize st}} = 
\boldsymbol{D}^{\mbox{\scriptsize st}}. 
\end{equation}
The breaking of the fluctuation-dissipation theorem 
has recently been recognized for the corresponding Langevin system\cite{SO}. 
\section{Path probability and detailed imbalance relation}
\label{sec3}
Following Tomita et al.\cite{TOT}, 
we apply the Onsager-Machlup theory\cite{OM,MO} to the Fokker-Planck equation, 
Eq.~(\ref{eq:Fokker_Planck_eq}). 
The time evolution of the probability distribution can be written 
in terms of the transition probability. 
\begin{equation}
p(\boldsymbol{x};t) = 
\int d\boldsymbol{x}' \; 
F\left (
\begin{array}{c|c}
\boldsymbol{x} & \boldsymbol{x}' \\
t & t_{0} 
\end{array}
\right ) 
p(\boldsymbol{x}';t_{0}).
\end{equation}
$F(\dots)$ is the transition probability. 
For a short-time propagation, 
the transition probability is evaluated as 
\begin{eqnarray}
& & 
F\left (
\begin{array}{c|c}
\boldsymbol{x}+ \Delta \boldsymbol{x} & \boldsymbol{x} \\
t+ \Delta t & t 
\end{array}
\right ) \nonumber \\
& = & 
\frac{1}{\sqrt{(2\pi)^{N}\det(\boldsymbol{D}(t))(\epsilon \Delta t)^{N}}} 
\nonumber \\
& & \times 
\exp \left [
-\frac{\Delta t \Omega}{4}
\left ( \frac{\Delta \boldsymbol{x}}{\Delta t}-\dot{\boldsymbol{y}}(\boldsymbol{x}) \right )^{t}
\cdot \boldsymbol{R}
\cdot 
\left ( \frac{\Delta \boldsymbol{x}}{\Delta t} -\dot{\boldsymbol{y}}(\boldsymbol{x}) \right )
\right ] 
+ {\cal O}((\Delta t)^{2}),
\label{eq:st_transition_prob}
\end{eqnarray}
where $\boldsymbol{R} = \boldsymbol{D}^{-1}$. 
From Eq.~(\ref{eq:st_transition_prob}), 
the Lagrangian for the path integral (i.e., the Onsager-Machlup function) 
is given by 
\begin{equation}
{\cal L}(\dot{\boldsymbol{x}},\boldsymbol{x})  = 
-\frac{\Omega}{4}
\left ( \dot{\boldsymbol{x}}-\dot{\boldsymbol{y}}(\boldsymbol{x}) \right )^{t}
\cdot \boldsymbol{R} \cdot 
\left ( \dot{\boldsymbol{x}}-\dot{\boldsymbol{y}}(\boldsymbol{x}) \right ).
\end{equation}
The path probability is given by 
\begin{eqnarray}
W_{\mbox{\scriptsize path}}(\{ \boldsymbol{x} \};A\rightarrow B) & = & 
\exp \left [\int_{t_{0}}^{t} ds \; {\cal L}(\dot{\boldsymbol{x}}^{*}(s),\boldsymbol{x}^{*}(s)) 
\right ],
\label{eq:path_prob}
\end{eqnarray}
where $\boldsymbol{x}^{*}(s)$ is to be taken along a given path $A\rightarrow B$. 
We set $A=\boldsymbol{x}(t_{0})$ and $B=\boldsymbol{x}(t)$. 
To calculate the path probability ratio, 
we evaluate the difference between the Lagrangians. 
\begin{equation}
{\cal L}(\dot{\boldsymbol{x}},\boldsymbol{x})  - {\cal L}(-\dot{\boldsymbol{x}},\boldsymbol{x})  
= \Omega \dot{\boldsymbol{x}}\cdot \boldsymbol{R} \cdot 
\dot{\boldsymbol{y}}(\boldsymbol{x}) .
\end{equation}
Here we assume that the probability distribution in the NESS 
is given by 
\begin{equation}
p^{\mbox{\scriptsize st}}(\boldsymbol{x}) \propto \exp[-\Omega \phi(\boldsymbol{x})].
\end{equation}
Thus, we have 
\begin{equation}
{\cal L}(\dot{\boldsymbol{x}},\boldsymbol{x})  - {\cal L}(-\dot{\boldsymbol{x}},\boldsymbol{x})  
= - \Omega \dot{\phi}(\boldsymbol{x})  + \Omega \dot{\boldsymbol{x}}\cdot 
\boldsymbol{R}^{\mbox{\scriptsize st}}
\cdot \left [ \dot{\boldsymbol{y}}(\boldsymbol{x}) 
+ \boldsymbol{D}^{\mbox{\scriptsize st}}\cdot \nabla \phi(\boldsymbol{x}) \right ], 
\end{equation}
in which we have used the relation 
$\dot{\phi}(\boldsymbol{x}) = \nabla \phi \cdot \dot{\boldsymbol{x}}$. 
Then for the NESS, we obtain 
\begin{equation}
\frac{P^{\mbox{\scriptsize st}}(A)W_{\mbox{\scriptsize path}}(\{ \boldsymbol{x} \};A\rightarrow B)}
{P^{\mbox{\scriptsize st}}(B)W_{\mbox{\scriptsize path}}(\{ \boldsymbol{x} \};B\rightarrow A)}
= 
\exp \left [ \Omega \int_{A}^{B} dt\; 
\dot{\boldsymbol{x}}^{t}\cdot \boldsymbol{R}^{\mbox{\scriptsize st}}\cdot 
\boldsymbol{v} \right ].
\label{eq:d_imb_rel}
\end{equation}
Similar relations were also obtained for Langevin systems 
by Taniguchi and Cohen,
\cite{TC1,TC2,TC3}\footnote{They called this relation the nonequilibrium detailed balance relation.} Seifert\cite{Seifert}, 
and Chernyak et al.\cite{CCJ}. 
If the detailed balance relation is satisfied, 
the right hand side of eq.~(\ref{eq:d_imb_rel}) is equal to $1$, 
i.e., it is in equilibrium and the entropy production is zero, 
because $\boldsymbol{\alpha}=0$.  
It is known that compared with the Onsager-Machlup theory, 
the argument of the exponential function is related to the entropy production. 
Therefore, the argument of the exponential function on the right-hand side 
is the entropy production rate for the path $A \rightarrow B$. 
Thus, we finally obtain a stochastic form of the entropy production rate, 
\begin{equation}
\sigma_{e}(\dot{\boldsymbol{x}},\boldsymbol{x}) = \Omega \, \dot
{\boldsymbol{x}}^{t}\cdot 
\boldsymbol{R}^{\mbox{\scriptsize st}}
\cdot \boldsymbol{v}(\boldsymbol{x}).
\label{eq:ep_term}
\end{equation}

Next, consider the average value of the entropy production term. 
As in the Onsager-Machlup theory, 
the most probable paths are categorized into two types, i.e., 
the forward evolution and the reverse evolution. 
For the forward evolution, 
the most probable path is given by 
\begin{equation}
\dot{\boldsymbol{x}} = \boldsymbol{K}^{\mbox{\scriptsize st}}\cdot \boldsymbol{x} + \boldsymbol{c} = 
\boldsymbol{K}^{\mbox{\scriptsize st}}\cdot (\boldsymbol{x}-\langle \boldsymbol{x} \rangle). 
\label{eq:most_prob_path1}
\end{equation}
Inserting Eq.~(\ref{eq:most_prob_path1}) into Eq.~(\ref{eq:ep_term}) gives
\begin{equation}
\sigma_{e} (\boldsymbol{x})= 
\Omega\,
\{ \boldsymbol{K}^{\mbox{\scriptsize st}}\cdot ( \boldsymbol{x} -  \langle \boldsymbol{x} \rangle ) \}^{t} 
\cdot \boldsymbol{R}^{\mbox{\scriptsize st}} 
\boldsymbol{\alpha}\cdot \boldsymbol{X}(\boldsymbol{x}).
\end{equation}
Taking an average over the NESS, we obtain
\begin{eqnarray}
\langle \sigma_{e} \rangle & = & 
\int d\boldsymbol{x} \; P^{\mbox{\scriptsize st}}(\boldsymbol{x}) \, \sigma_{e} (\boldsymbol{x}) \nonumber \\
& = & 
-\mbox{Tr}(\boldsymbol{\alpha} \boldsymbol{R}^{\mbox{\scriptsize st}} \boldsymbol{\alpha} \boldsymbol{g}^{\mbox{\scriptsize st}}).
\label{eq:ep_central}
\end{eqnarray}
This is the central result of this section, 
i.e., another form of the detailed imbalance expression. 
Note that the entropy production is expressed 
in terms of the irreversible circulation $\boldsymbol{\alpha}$. 
If in equilibrium, i.e., the detailed balance is satisfied, 
$\boldsymbol{\alpha}$ is zero. Then, the entropy production vanishes. 
This is consistent with the physical requirement. 
It is important that the entropy production is expressed 
as a quadratic form of $\boldsymbol{\alpha}$. 
\section{Examples}\label{sec4}
In this section, 
we check whether or not the derived expression of the entropy production 
coincides with the thermodynamical expression. 
Two examples are considered. 
One is a chemical reaction network. 
The other is a one-dimensional diffusion system. 
\subsection{Chemical reaction network} \label{sec4-1}
Let us consider the following simple case: 
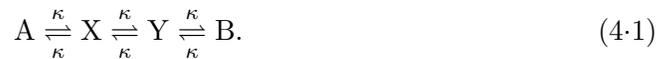
\begin{equation}
{\rm A} \reaction{\kappa}{\kappa} {\rm X} \reaction{\kappa}{\kappa} {\rm Y} 
\reaction{\kappa}{\kappa} {\rm B}.
\end{equation}
Here all rate constants are equal to $\kappa$. 
To maintain the steady state, 
the concentrations of the chemical species ${\rm A}$ and ${\rm B}$ are kept 
constant by reservoirs. 
This chemical reaction system is linear. 
Using Eqs.~(\ref{eq:def_K}) and (\ref{eq:c1}), 
the matrix $\boldsymbol{K}^{\mbox{\scriptsize st}}$ and the vector $\boldsymbol{c}$ are given by 
\begin{equation}
\boldsymbol{K}^{\mbox{\scriptsize st}} = \kappa
\left (
\begin{array}{cc}
-2 & 1 \\
1 & -2
\end{array}
\right ),\; 
\boldsymbol{c} = \kappa 
\left (
\begin{array}{c}
\langle a \rangle\\
\langle b \rangle
\end{array}
\right ),
\end{equation}
where $\langle a \rangle =\langle A\rangle /\Omega$ 
and $\langle b \rangle =\langle B\rangle/\Omega$. 
Using Eq.~(\ref{eq:def_D}) for $\boldsymbol{D}^{\mbox{\scriptsize st}}$, 
Eq.~(\ref{eq:def_K}) for $\boldsymbol{K}^{\mbox{\scriptsize st}}$, 
and making the right-hand side of Eq.~(\ref{eq:time_evol_sigma}) 
equal to zero,  
we obtain the matrix $\boldsymbol{\sigma}^{\mbox{\scriptsize st}}$. 
\begin{equation}
\boldsymbol{\sigma}^{\mbox{\scriptsize st}} = 
\frac{1}{3}
\left (
\begin{array}{cc}
2\langle a\rangle + \langle b \rangle& 0 \\
0 & \langle a\rangle+2 \langle b \rangle 
\end{array}
\right ).
\end{equation}
Using Eq.~(\ref{eq:def_alpha}), 
the matrix $\boldsymbol{\alpha}$ is given by 
\begin{equation}
\boldsymbol{\alpha} = 
\frac{\kappa(\langle a \rangle- \langle b \rangle)}{6}
\left (
\begin{array}{cc}
0 & 1 \\
-1 & 0 
\end{array}
\right ).
\end{equation}
The entropy production derived from the Fokker-Planck equation becomes 
\begin{eqnarray}
\langle \sigma_{e} \rangle 
&=& - \mbox{Tr}(\boldsymbol{\alpha} \boldsymbol{R}^{\mbox{\scriptsize st}} \boldsymbol{\alpha} 
\boldsymbol{g}^{\mbox{\scriptsize st}}) \nonumber \\
& = & \kappa \frac{4(\langle a \rangle -\langle b \rangle)^{2}}
{23\langle a \rangle^{2}+62\langle a\rangle \langle b \rangle 
+ 23\langle b \rangle ^{2}} \nonumber \\
& \approx & \frac{4\kappa(\langle a \rangle -\langle b \rangle)^{2}}
{27(\langle a \rangle + \langle b \rangle)^{2}}.
\label{eq:ep_chem_Fokker}
\end{eqnarray}
The last line is the approximation near equilibrium, 
i.e., $\langle a \rangle \sim \langle b \rangle$. 
We have employed an expansion with a symmetric form. 

In the thermodynamical consideration, the entropy production is given by 
\begin{eqnarray}
\sigma_{e,\mbox{\scriptsize th}} 
& = & \sum_{i=1}^{3} {\cal J}_{i} \frac{{\cal A}_{i}}{T}, 
\end{eqnarray}
where
\begin{eqnarray}
{\cal J}_{1} = \kappa (\langle A \rangle-\langle X \rangle), &\hspace*{1cm}& 
{\cal A}_{1} = T \log\frac{\langle A \rangle}{\langle X \rangle}, \\
{\cal J}_{2} = \kappa (\langle X \rangle-\langle Y \rangle), &\hspace*{1cm}& 
{\cal A}_{2} = T \log\frac{\langle X \rangle}{\langle Y \rangle}, \\
{\cal J}_{3} = \kappa (\langle Y \rangle-\langle B \rangle), &\hspace*{1cm}& 
{\cal A}_{3} = T \log\frac{\langle Y \rangle}{\langle B \rangle}.
\end{eqnarray}
${\cal J}_{i}$ is the reaction rate of reaction $i$ and 
${\cal A}_{i}$ is the affinity of reaction $i$. 
This expression is equivalent to Eq.~(\ref{eq:ep_Gaspard_exact}).
Near equilibrium, the entropy production is 
\begin{equation}
\sigma_{e,\mbox{\scriptsize th}} 
\approx 
\frac{2\kappa \Omega (\langle a \rangle - \langle b \rangle)^{2}}
{3(\langle a \rangle + \langle b \rangle)}.
\label{eq:ep_chem_thermo}
\end{equation}
The result of Eq.~(\ref{eq:ep_chem_Fokker}) disagrees 
with the thermodynamical result of Eq.~(\ref{eq:ep_chem_thermo}). 
\subsection{One-dimensional diffusion system} \label{sec4-2}
In this subsection, a one-dimensional diffusion system is considered. 
The system is a pipe with the cross section $\Sigma$. 
This pipe is divided into $L$ cells of length $\lambda$. 
Thus, the volume of each cell is $\Omega = \lambda \Sigma$. 
Particles exhibit a random walk between cells. 
All rate constants are given by $\kappa$. 
The reaction is represented as 
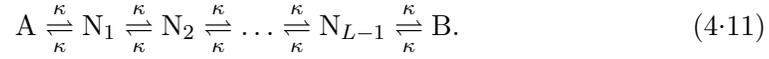
\begin{equation}
{\rm A} \reaction{\kappa}{\kappa} 
{\rm N}_{1} \reaction{\kappa}{\kappa}  
{\rm N}_{2} \reaction{\kappa}{\kappa} 
\dots \reaction{\kappa}{\kappa} 
{\rm N}_{L-1} \reaction{\kappa}{\kappa} {\rm B}.
\end{equation}
The time evolution of the population $N_{i}$ is determined 
by the following rate equation:
\begin{equation}
\dot{N}_{i} = \kappa (N_{i+1}-2N_{i} + N_{i-1}),
\end{equation}
and at both edges,
\begin{equation}
N_{0} = A,\; N_{L}=B. 
\end{equation}
At the edges, a constant number of particles is supplied by reservoirs. 
We denote the density of the particle in the $i$th cell by $n_{i}$, 
i.e., $n_{i} = N_{i}/\Omega$. 
\begin{eqnarray}
\dot{n}_{i} & = & \kappa (n_{i+1}-2n_{i} + n_{i-1}) \nonumber \\
& \approx & 
\kappa \lambda^{2} \nabla^{2} n.
\end{eqnarray}
Thus, the spatial diffusion coefficient is given by 
\begin{equation}
{\cal D} = \kappa \lambda^{2}. 
\end{equation}
Now let us consider the master equation for this system. 
The transition probabilities are given by
\begin{eqnarray}
W(\dots,N_{i},N_{i+1},\dots \rightarrow \dots,N_{i}-1,N_{i+1}+1,\dots) 
& = & \kappa N_{i}, \\
W(\dots,N_{i},N_{i+1},\dots \rightarrow \dots,N_{i}+1,N_{i+1}-1,\dots) 
& = & \kappa N_{i+1}. 
\end{eqnarray}
This problem was analyzed in the context of the fluctuation theorem\cite{AP}. 
The steady solution of the master equation is multi-Poissonian.
\begin{equation}
P^{\mbox{\scriptsize st}}(N_{1},N_{2},\dots,N_{L-1}) = 
\prod_{i=1}^{L-1}e^{-\langle N_{i} \rangle} 
\frac{\langle N_{i} \rangle ^{N_{i}}}{N_{i}!}. 
\label{eq:m_Poissonian}
\end{equation}
The thermodynamical entropy production is given by 
\begin{eqnarray}
\sigma_{e,\mbox{\scriptsize th}} & = & 
\frac{\kappa (\langle A \rangle - \langle B \rangle)}{L} 
\log \frac{\langle A \rangle}{ \langle B \rangle}.
\label{eq:ep_diffusion_chem}
\end{eqnarray}
This equation is equivalent to the Schnakenberg-Gaspard expression, 
Eq.~(\ref{eq:ep_Gaspard_exact}). 
One can confirm by inserting Eq.~(\ref{eq:m_Poissonian}) 
into Eq.~(\ref{eq:ep_Gaspard_exact}), 
that Eq.~(\ref{eq:ep_diffusion_chem}) is satisfied. 
In the continuous limit, we have
\begin{eqnarray}
\sigma_{e,\mbox{\scriptsize th}} & = & 
\Sigma {\cal D} \int_{0}^{L\lambda} dx \frac{|\nabla n(x)|^{2}}{n(x)},
\label{eq:ep_diffusion_n_equi}
\end{eqnarray}
where $\Sigma \; dx$ is the volume element here. 
In the last line, we employed the linear approximation near equilibrium. 

Using Eq.~(\ref{eq:moments}), the moments are calculated as 
\begin{eqnarray}
\boldsymbol{c}_{1,i} & = & \kappa(n_{i+1}-2n_{i}+n_{i-1}), \\
\boldsymbol{c}_{2,i,i} & = & \kappa(n_{i+1}+2n_{i}+n_{i-1}), \\
\boldsymbol{c}_{2,i,i+1} & = & \boldsymbol{c}_{2,i+1,i} = \kappa(n_{i}+n_{i+1}).
\end{eqnarray}
For $\boldsymbol{c}_{2}$, other entries are zero. 
The matrices $\boldsymbol{K}^{\mbox{\scriptsize st}}$ and $\boldsymbol{D}^{\mbox{\scriptsize st}}$ are given by 
\begin{equation}
\boldsymbol{K}^{\mbox{\scriptsize st}} = \kappa \left (
\begin{array}{ccccccc}
 -2 & 1  & 0  & \cdots &\dots & 0\\
 1 & -2 & 1 & \cdots & \cdots& 0\\
0 & 1 & -2 & \ddots & \cdots& 0\\
 \vdots &  \cdots & \ddots & \ddots & \ddots  & \vdots\\
 0  &  \cdots  & \cdots  & 1 &  -2 & 1\\
 0  & \cdots  & \cdots& 0 & 1 &  -2\\
\end{array}
\right )
\end{equation}
and
\begin{eqnarray}
D^{\mbox{\scriptsize st}}_{i,i} & = & 
\frac{\kappa}{2}(n_{i-1}^{\mbox{\scriptsize st}}+
2n_{i}^{\mbox{\scriptsize st}}+n_{i+1}^{\mbox{\scriptsize st}}) 
= 2 \kappa n_{i}^{\mbox{\scriptsize st}}, \\
D^{\mbox{\scriptsize st}}_{i,i+1} & = & 
D^{\mbox{\scriptsize st}}_{i+1,i} = 
-\frac{\kappa}{2}(n_{i}^{\mbox{\scriptsize st}}+n_{i+1}^{\mbox{\scriptsize st}}). 
\end{eqnarray}
For $\boldsymbol{D}^{\mbox{\scriptsize st}}$, other entries are zero. 
We used the fact that the steady solution is given by 
\begin{equation}
n_{i}^{\mbox{\scriptsize st}} = n_{0}- i\lambda |\nabla n|.
\end{equation}
The variance matrix $\boldsymbol{\sigma}^{\mbox{\scriptsize st}}$ 
and the circulation matrix $\boldsymbol{\alpha}$ are given by 
\begin{eqnarray}
\boldsymbol{\sigma}^{\mbox{\scriptsize st}} & =&  
\left (
\begin{array}{cccccc}
 n_{1}^{\mbox{\scriptsize st}} & 0    & 0  &    & \dots & 0\\
 0    & n_{2}^{\mbox{\scriptsize st}} & 0  &    & \dots & 0\\
 \vdots  &      & \ddots   &    &  & \vdots\\
 0    & \dots &    &   &  n_{L-2}^{\mbox{\scriptsize st}}  & 0\\
 0    & \dots &    &    & 0     & n_{L-1}^{\mbox{\scriptsize st}}\\
\end{array}
\right )
\end{eqnarray}
and 
\begin{eqnarray}
\boldsymbol{\alpha} & = & 
\frac{1}{2} (\boldsymbol{\sigma}^{\mbox{\scriptsize st}}\widetilde{\boldsymbol{K}}^{\mbox{\scriptsize st}} 
- \boldsymbol{K}^{\mbox{\scriptsize st}}\boldsymbol{\sigma}^{\mbox{\scriptsize st}}) 
\nonumber \\
& = & 
\frac{\kappa \lambda \nabla n}{2}
\left (
\begin{array}{ccccccc}
 0 & 1  & 0  & \cdots &\dots & 0\\
 -1 & 0 & 1 & \cdots & \cdots& 0\\
0 & -1 & 0 & \ddots & \cdots& 0\\
 \vdots &  \cdots & \ddots & \ddots & \ddots  & \vdots\\
 0  &  \cdots  & \cdots  & -1 &  0 & 1\\
 0  & \cdots  & \cdots& 0 & -1 &  0\\
\end{array}
\right ).
\end{eqnarray}
The entropy production for the corresponding Fokker-Planck equation 
is given by Eq.~(\ref{eq:ep_central}). Then we have
\begin{eqnarray}
\langle \sigma_{e} \rangle & =&  
- \mbox{Tr}(\boldsymbol{\alpha}\boldsymbol{R}^{\mbox{\scriptsize st}}
\boldsymbol{\alpha}\boldsymbol{g}^{\mbox{\scriptsize st}})
\nonumber \\
& \approx & 
\frac{\kappa\lambda^{2}|\nabla n|^{2}}{4\overline{n}^{2}}
\frac{2(L-2)(L-1)}{L} \nonumber \\
& = & 
\frac{{\cal D}}{2\lambda \overline{n}} \int_{0}^{L\lambda} dx\; 
\frac{|\nabla n(x)|^{2}}{n(x)}. \mbox{\hspace*{1cm}} (L \rightarrow \infty)
\label{eq:ep_central1}
\end{eqnarray}
This result does not agree with 
the thermodynamical result of  
Eq.~(\ref{eq:ep_diffusion_n_equi}). 
In particular, the order of Eq.~(\ref{eq:ep_central1}) is different 
by a factor of $1/\Omega$ compared with that of 
Eq.~(\ref{eq:ep_diffusion_n_equi}) and 
the concentration dependence disagrees.  

\section{Path weight principle} \label{sec5}
As shown in the previous section, the entropy production derived 
directly from our Fokker-Planck equation disagrees with that of 
the original master equation and the thermodynamical equation.  
This discrepancy should be examined. 

First, let us consider the reason for this discrepancy. 
Our original master equation describes the phenomena of discrete 
jumps such as the occasional collisions in the chemical reaction system. 
On the other hand, the corresponding Fokker-Planck equation treats the 
averaged continuous evolution of the original physical random process. 
This relation is similar to that between a random walk and 
Brownian motion, which corresponds to the former. Note that the different
 random walks, say, 
(i) random jump $\pm \Delta$ at each mean interval $\tau_{0}$, 
and (ii) random jump $\pm 2 \Delta$ at each mean interval $4\tau_{0}$, 
are described by the same Brownian motion with the diffusion coefficient 
$D=\Delta^{2}/2\tau_{0}$. However, the entropy production differs for 
each case, i.e., that of case (i) is four times larger 
than that of case (ii). 
This fact tells us that the Fokker-Planck equation and 
the Brownian motion cannot be used for the purpose of 
calculating the correct entropy production, at least,
when the original master equation describes a discrete stochastic process. 
However, they accurately describe the long-time evolution of 
the probability itself due to the central limit theorem.
The entropy production is due to the short-time 
behavior of fluctuations, i.e., the detailed imbalance relation. 
Entropy is created at each discrete jump process such as reactive collisions 
among atoms or molecules in the chemical reaction.
Let us call these discrete jump process 
the elementary process of entropy production.

Second, let us introduce the {\it path weight principle}, which is a 
type of correspondence rule for overcoming this difficulty. 
The above consideration suggests that 
to calculate the correct entropy production 
in the present Fokker-Planck scheme, 
we should take account of the number of elementary processes
included in a given continuous stochastic path.

For example, let us consider the following chemical reaction network, 
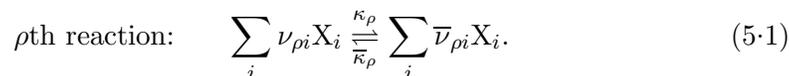
\begin{equation}
\mbox{$\rho$th reaction: \hspace*{0.5cm}} 
\sum_{i} \nu_{\rho i} {\rm X}_{i} 
\reaction{\kappa_{\rho}}{\overline{\kappa}_{\rho}}
\sum_{i} \overline{\nu}_{\rho i} {\rm X}_{i}. 
\end{equation}
The elementary random walk comprises each reactive collision which 
means that $\Delta Y_{\rho} = \pm 1$ in a mean interval $\kappa_{\rho}^{-1}
(\overline{\kappa}_{\rho}^{-1})$, where $Y_{\rho}$ is the reaction coordinate
of the $\rho$th reaction defined by 
\begin{equation}
\delta X_{i} = 
\sum_{\rho}(\overline{\nu}_{\rho i} - \nu_{\rho i}) \delta Y_{\rho}. 
\end{equation}
Using the reaction coordinates as the set of stochastic variables, 
the frequencies of the positive and negative reactions 
in the $\rho$th direction per unit time 
are related to the second moment of the transition probability, 
\begin{equation}
2 \Omega {D'}^{\mbox{\scriptsize st}}_{\rho}
= W(\Delta Y_{\rho} = + 1)(+1)^{2} + 
W(\Delta Y_{\rho} = - 1)(-1)^{2},
\end{equation}
when the reaction flow can be neglected in near-equilibrium situations.
This condition will be satisfied in the linearized, local-equilibrium
estimation below.

Thus, the diffusion constants directly give the number of elementary random
walks in a unit time if the reaction rates satisfy $\kappa_{\rho}=
\overline{\kappa}_\rho$.
However, it is difficult to find a general correspondence rule for the
population coordinate $\{X_{i}\}$ except for the following special
cases.

\subsection{One-dimensional diffusion system} \label{sec5-1}
In the diffusion model used in \S~\ref{sec4}, 
the rate constants $\kappa_{\rho}$'s are assumed to be a constant, 
$\kappa$, i.e.,
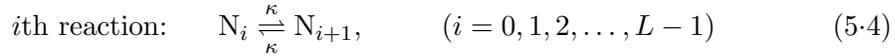
\begin{equation}
\mbox{$i$th reaction: \hspace*{0.5cm}} {\rm N}_{i} \reaction{\kappa}{\kappa} {\rm N}_{i+1},\mbox{\hspace*{1cm}} (i=0,1,2,\dots,L-1)
\end{equation}
where ${\rm N}_0={\rm A}$ and ${\rm N}_{L}={\rm B}$.
The diffusion matrix in the reaction coordinates is given by a diagonal matrix
$\{ D'^{\mbox{\scriptsize st}}_{i} \delta_{ij}\}$, where
\begin{equation}
{D'}^{\mbox{\scriptsize st}}_{i}=
\frac{\kappa}{2}(n^{\mbox{\scriptsize st}}_{i}+n^{\mbox{\scriptsize st}}_{i+1}),
\end{equation}
with $n_i=N_i/\Omega$.
The elementary process in this case is a jump of one particle in a given
cell to the left or right cell in the mean interval time, $\kappa^{-1}$,
that is, a uniform random walk in the one-dimensional real space.
Therefore, at least in this special case, the number of elementary
jump processes included in a continuous unit-time path in the $i$th direction
is given by the ratio, $2\Omega {D'}^{\mbox{\scriptsize st}}_{i}/\kappa$.
Then the correspondence rule in this case is given by
\begin{equation}
\dot{\eta}_{i} \longrightarrow 
2\kappa^{-1}\Omega{D'}^{\mbox{\scriptsize st}}_{i} \dot{\eta}_{i}, 
\label{eq:PWP1}
\end{equation}
where $\eta_{\rho}=Y_{\rho}/\Omega$.
Thus, the diffusion constant can be used for the path weight.

In the concentration space $\{ n_{i} \}$, 
let us assume that the same correspondence
rule can be applied in the principal-axis space where the diffusion
matrix $\{D^{\mbox{\scriptsize st}}_{ij}\}$ is diagonalized, although we have
no definite principle for determining the coefficient $\kappa^{-1}$ itself in
the present case.
Note that this space is not necessarily equivalent 
to the reaction coordinate space $\{\eta_i\}$.
We have the linear relations, 
\begin{equation}
\dot{n}_{i} = \sum_{\rho}(\delta_{\rho,i-1} - \delta_{\rho,i})
\dot{\eta}_{\rho},
\end{equation}
and
\begin{equation}
D_{ij}^{\mbox{\scriptsize st}} = \sum_{\rho} (\delta_{\rho,i-1} - \delta_{\rho,i})
(\delta_{\rho,j-1} - \delta_{\rho,j}){D'}^{\mbox{\scriptsize st}}_{\rho},
\end{equation}
where $\delta_{\rho,i}$ is the usual Kronecker delta. 
However, this transformation is not represented as a square matrix, 
i.e., it is not invertible and not an orthogonal transformation. 

Thus, we obtain a {\em hypothetical} correspondence rule 
in the concentration space
\begin{equation}
\dot{n}_{i} \longrightarrow 
\sum_{j}2\kappa^{-1}\Omega{D}^{\mbox{\scriptsize st}}_{ij}\dot{n}_{j}.  
\label{eq:PWP2}
\end{equation}
Let us call this the {\em path weight principle}.

As a result, we have a corrected expression for the entropy production
rate,
\begin{equation}
\langle \sigma_{e} \rangle '
=-2\kappa^{-1}\Omega~
\mbox{Tr}( \boldsymbol{\alpha} \boldsymbol{g}^{\mbox{\scriptsize st}}
\boldsymbol{\alpha} ). 
\end{equation}
Here the prime means that the entropy production is modified 
by the path weight principle. 
Using the explicit forms of $\boldsymbol{\alpha}$ and 
$\boldsymbol{g}^{\mbox{\scriptsize st}}$ in \S~\ref{sec4}, 
we obtain a final result,
\begin{equation}
\langle \sigma_{e} \rangle '
=\Sigma \int_{0}^{L\lambda} \frac{{\cal D}|\nabla n(x)|^2}{n(x)} dx,
\end{equation}
in the continuum limit $L\rightarrow \infty$, where ${\cal D}=\kappa\lambda^2$
is the spatial diffusion constant.
This coincides exactly with the thermodynamic result, 
i.e., Eq.~(\ref{eq:ep_diffusion_n_equi}).
\subsection{Chemical reaction network} \label{sec5-2}
The path weight principle may also be applied to the simple chemical reaction 
($L=3$) in \S~\ref{sec4-1}.
An simple result for the evaluation of the entropy production is given by
\begin{equation}
\langle \sigma_{e} \rangle '
=\frac{2\kappa\Omega (\langle a\rangle-\langle b\rangle)^2}{9(\langle a\rangle+
\langle b\rangle)},
\end{equation}
in a symmetrized form near equilibrium. Compared with
Eq.~(\ref{eq:ep_chem_thermo}), there is a difference of a factor of 1/3.

The reason for this is evident.
We should treat the boundary effect more carefully for finite $L$.
In the present case there are three different random walks 
in the $xy$-plane, i.e., 
two boundary modes $\dot{Y}_0$ along the $x$-axis and $\dot{Y}_2$
along the $y$-axis in addition to the diagonal mode $\dot{Y}_1$ in 
the direction $(1,-1)$.
Therefore, the elementary processes are not isotropic in the $xy$-plane.

Instead of performing this ambiguous transformation, 
the reason for the factor of 1/3 may be that
the irreversible circulations 
corresponding to both end reactions have not been taken into account 
in the present scheme. 
It can be easily shown that this factor of 1/3 is removed 
when the variables $A$ and $B$ are added to the set of stochastic variables 
as 
\begin{eqnarray}
\dot{A} & = & \kappa (X-A) - c, \\
\dot{X} & = & \kappa (A-2X+Y), \\
\dot{Y} & = & \kappa (X-2Y+B), \\
\dot{B} & = & \kappa (Y-B)- c', 
\end{eqnarray}
where $c$ and $c'$ are parameters controlled by external equilibrium 
reservoirs to keep $A$ and $B$ constant. 
Here $c=\kappa( \langle X \rangle - A) $ and 
$c'= \kappa ( \langle Y \rangle - B)$.
Except for the matrix $\boldsymbol{K}^{\mbox{\scriptsize st}}$, 
which is modified slightly to 
\begin{equation}
\boldsymbol{K}^{\mbox{\scriptsize st}}
=\kappa
\left (
\begin{array}{cccc}
-1 & 1  & 0  &  0\\
1 & -2 & 1  &  0\\
0 & 1  & -2 &  1\\
 0 & 0 & 1   & -1\\
\end{array}
\right ),
\end{equation}
the other quantities are exactly the same as those given 
in \S~\ref{sec4-2} for $L-1=4$.

\section{Concluding remarks} \label{sec6}
We have shown that 
the entropy production of the Fokker-Planck 
equation derived from the master equation differs from 
that for the original master equation. 
The reason for this is clearly due to the fact that 
the master equation treats discrete events, 
but the Fokker-Planck equation is an approximation of the master equation. 
In the Fokker-Planck equation, the original discrete events are smoothed out. 
To evaluate the entropy production, 
one has to recover the discreteness of the events 
in the treatment of the corresponding Fokker-Planck equation. 
To overcome this problem, 
we have proposed the {\it path weight principle}. 
The entropy production from the corresponding Fokker-Planck equation 
is modified by multiplying by the diffusion coefficient. 
For two simple examples, it has been demonstrated 
that the path weight principle yields 
the entropy production for the original master equation. 

At present, we do not know whether or not the path weight principle 
can be applied to any type of master equation. 
However, we believe that 
the path weight principle can be applied, at least, to the cases 
in which jumps in the transitions are small compared with $\Omega$, 
namely $|\Delta| \sim 1$. 
\section*{Acknowledgements}
The authors are grateful to Professor K.~Kitahara 
for enlightening discussions and continuous encouragement.

%
%



\end{document}